\documentclass[twocolumn,showpacs,preprintnumbers,amsmath,amssymb,floatfix,a4paper]{revtex4}

\usepackage{graphicx}
\usepackage{dcolumn}
\usepackage{bm}
\usepackage{setspace}
\usepackage{amsmath}

\begin{document}


\title{Thermonuclear $^{42}$Ti($p$,$\gamma$)$^{43}$V rate in type I X-ray bursts}

\author{J.J.~He$^1$}
\email{jianjunhe@impcas.ac.cn}
\author{A. Parikh$^{2,3}$}
\author{B.A. Brown$^{4}$}
\author{T. Rauscher$^{5,6}$}
\author{S.Q. Hou$^{1,7}$}
\author{Y.H. Zhang$^{1}$}
\email{yhzhang@impcas.ac.cn}
\author{X.H. Zhou$^{1}$}
\author{H.S. Xu$^{1}$}

\affiliation{
$^{1}$Key Laboratory of High Precision Nuclear Spectroscopy and Center for Nuclear Matter Science, Institute of Modern Physics, Chinese Academy of Sciences, Lanzhou 730000, China\\
$^{2}$Departament de F\'{\i}sica i Enginyeria Nuclear, EUETIB, Universitat Polit\`{e}cnica de Catalunya, Barcelona E-08036, Spain\\
$^{3}$Institut d'Estudis Espacials de Catalunya, Barcelona E-08034, Spain\\
$^{4}$Department of Physics and Astronomy and National Superconducting Cyclotron Laboratory, Michigan State University, East Lansing, Michigan 48824, USA\\
$^{5}$Centre for Astrophysics Research, School of Physics, Astronomy and Mathematics, University of Hertfordshire, Hatfield AL10 9AB, United Kingdom\\
$^{6}$Departement f\"{u}r Physik und Astronomie, Universit\"{a}t Basel, Klingelbergstrasse 82, Basel CH-4056, Switzerland\\
$^{7}$University of Chinese Academy of Sciences, Beijing 100049, China
}

\date{\today}  

\begin{abstract}
The thermonuclear rate of the $^{42}$Ti($p$,$\gamma$)$^{43}$V reaction has been reevaluated based on a recent precise proton separation
energy measurement of $S_p$($^{43}$V)=83$\pm$43 keV. The astrophysical impact of our new rates has been
investigated through one-zone postprocessing type I x-ray burst calculations. It shows that the new experimental value of $S_p$
significantly affects the yields of species between A$\approx$40--45. As well, the precision of the recent experimental $S_p$ value
constrains these yields to better than a factor of three.
\end{abstract}



\maketitle

\section{Introduction}
Type I X-ray bursts (XRBs) arise from thermonuclear runaways within the accreted envelopes of neutron stars in close binary systems~\cite{bib:woo76,bib:jos77}.
About one hundred bursting systems have been identified in the Galaxy, with light curves of about 10--100 s in duration, recurrence
periods of $\sim$ hours to days, and peak luminosity L$_\mathrm{peak}$$\approx$10$^{4}$--10$^{5}$ L$_\mathrm{\odot}$ (similar,
{\it e.g.}, to L$_\mathrm{peak}$ of classical novae). During the thermonuclear runaway, an accreted envelope enriched in H and He may be
transformed to matter strongly enriched in heavier species (up to A$\approx$100~\cite{bib:sch01,bib:elo09}) via the $\alpha$p-process and
the rapid proton capture process (rp-process)~\cite{bib:wal81,bib:sch98,bib:woo04}. Current
XRB models do not predict the ejection of any appreciable amounts of synthesized material during the burst. Nonetheless, calculations
indicate that radiative winds generated during some bursts may eject material. Studies are ongoing to examine the viability of detecting any
associated absorption features. For reviews on aspects of type I X-ray bursts, see, {\it e.g.}, Refs.~\cite{bib:lew93,bib:str06,bib:par13}.

The rp-process is largely characterized by localized $({\rm p},\gamma)$-$(\gamma,{\rm p})$ equilibrium within particular isotonic
chains near the proton drip-line. Slower $\beta$-decays (followed by fast ($p$,$\gamma$) reactions) connect these isotonic chains
and set the timescale for processing towards heavier nuclei. In such an equilibrium situation the abundance distribution within an
isotonic chain depends exponentially on nuclear mass differences as the abundance ratio between two neighboring isotones is proportional
to $\exp[S_p/kT]$, where $S_p$ is the proton separation energy and $T$ the temperature of the stellar environment.
In particular, those isotonic chains with sufficiently small $S_p$ values (relative to XRB temperatures - at 1 GK, $kT$$\approx$100 keV) need to be
known with a precision of at least 50--100 keV~\cite{bib:sch98,bib:par09}. These include, among others, $S_p$($^{26}$P), $S_p$($^{43}$V),
$S_p$($^{46,47}$Mn), $S_p$($^{61}$Ga), and $S_p$($^{65}$As)~\cite{bib:par09}. As well, reliable nuclear physics input (including precise mass
values and nuclear structure information) is needed for those nuclei along the rp-process path to calculate the thermonuclear reaction rates
required for XRB models. Model predictions can then be compared with {\it e.g.}, observations of XRB light curves to extract quantitative
information about the stellar environments~\cite{bib:sch06}.

The level structure of $^{43}$V is not experimentally known. The thermonuclear rate of the $^{42}$Ti($p$,$\gamma$)$^{43}$V
reaction was first estimated by Wormer {\it et al.}~\cite{bib:wor94} based entirely on the properties of four states in the mirror
nucleus $^{43}$Ca~\cite{bib:end78,bib:end90}. Later, this rate was recalculated by Herndl {\it et al.}~\cite{bib:her95} using two
states determined through a shell model calculation of $^{43}$V. Theoretical rates calculated using statistical models are
available~\cite{bib:jina}; however, due to the low density of excited states expected in $^{43}$V near the proton threshold, such
calculations are not ideal for this reaction~\cite{bib:rau97,bib:rau00,bib:rau01}. A theoretical value of $S_p$=90$\pm$200 keV from
the atomic mass evaluation (AME85)~\cite{bib:aud85} was utilized in the above rate calculations. Another theoretical value of
$S_p$=190$\pm$230 keV was adopted in later AME95~\cite{bib:aud95} and AME03~\cite{bib:aud03} compilations.

Recently, precise mass measurements of nuclei along the rp-process path have become available. These measurements were made at the
HIRFL-CSR (Cooler-Storage Ring at the Heavy Ion Research Facility in Lanzhou)~\cite{bib:xia02} in an IMS (Isochronous Mass Spectrometry)
mode. Masses measured include those of a series of $T_z$=-1/2 nuclei ($^{63}$Ge, $^{65}$As, $^{67}$Se, and
$^{71}$Kr)~\cite{bib:tu11,bib:zhang12} and $T_z$=-3/2 nuclei ($^{41}$Ti, $^{43}$V,
$^{45}$Cr, $^{47}$Mn, $^{49}$Fe, $^{53}$Ni, and $^{55}$Cu)~\cite{bib:yan13}. The proton separation energy of $^{43}$V
has been experimentally determined to be $S_p$=83$\pm$43 keV for the first time~\cite{bib:yan13}. Although the predicted values in
the previous compilations (AME85, AME95 and AME03) agree with the experimental value within 1~$\sigma$ uncertainties, the latter
is significantly more precise. This allows the uncertainty in the rate of the $^{42}$Ti($p$,$\gamma$)$^{43}$V reaction to be
dramatically reduced. In this work, the thermonuclear rate of $^{42}$Ti($p$,$\gamma$)$^{43}$V has been reevaluated using the recent
experimental $S_p$ ($^{43}$V) value and new calculated resonant and direct capture (DC) rates. The astrophysical impact of our
new rates has been investigated through one-zone postprocessing x-ray burst calculations.

\section{Reaction rate calculation}
\subsection{Resonant rate}
We begin by estimating the $^{42}$Ti($p$,$\gamma$)$^{43}$V resonant rate using exactly the level energies, half-lives and single-particle
spectroscopic factors from the mirror nucleus $^{43}$Ca~\cite{bib:end78}. A similar approach was used in Ref.~\cite{bib:wor94}.
The resonant rate is calculated by the well-known narrow resonance formalism~\cite{bib:wor94,bib:her95,bib:rol88},
\begin{widetext}
\begin{eqnarray}
N_A\langle \sigma v \rangle_\mathrm{res}=1.54 \times 10^{11} (AT_9)^{-3/2} \omega\gamma \mathrm{[MeV]} \mathrm{exp} \left (-\frac{11.605E_r \mathrm{[MeV]}}{T_9} \right) [\mathrm{cm^3s^{-1}mol^{-1}}].
\label{eq1}
\end{eqnarray}
\end{widetext}
Here, the resonant energy $E_r$ and strength $\omega\gamma$ are in units of MeV. For the proton capture reaction, the reduced mass
$A$ is defined by $A_T$/(1+$A_T$) where $A_T$ is the target mass. The resonant strength $\omega\gamma$ is defined by
\begin{eqnarray}
\omega\gamma=\frac{2J+1}{2(2J+1)}\frac{\Gamma_p\times\Gamma_\gamma}{\Gamma_\mathrm{tot}}.
\label{eq2}
\end{eqnarray}
Here, $J_T$ and $J$ are the spins of the target and resonant state, respectively. $\Gamma_p$ is the partial width for the entrance
channel, and $\Gamma_\gamma$ is that for the exit channel. In the excitation energy range considered in this work, other decay
channels are closed~\cite{bib:aud03}, and hence the total width $\Gamma_\mathrm{tot}$$\approx$$\Gamma_p$+$\Gamma_\gamma$.
Similar to the approach used by Wormer {\it et al.}, the gamma partial widths of the unbound states in $^{43}$V were estimated by
the life-times ($\tau$) of the corresponding bound states in the mirror $^{43}$Ca via $\Gamma_\gamma$=$\hbar$/$\tau$; the proton partial widths were
calculated by the following equation,
\begin{eqnarray}
\Gamma_{p}=\frac{3\hbar^2}{AR^2}P_{\ell}(E)C^2S_p.
\label{eq3}
\end{eqnarray}
Here, $R$=1.26$\times$(1+42$^{\frac{1}{3}}$) fm is the nuclear channel radius~\cite{bib:wor94}, $P_\ell$ the Coulomb penetrability
factor, and $C^2S_p$ the proton spectroscopic factor of the resonance.

For this reaction, a temperature of 2 GK corresponds to a Gamow peak $E_{x}$($^{43}$V)$\approx$1.5 MeV with a width of
$\Delta$$\approx$1.2 MeV~\cite{bib:rol88}. Therefore, its resonant rate is determined by the excited states of $^{43}$V up to
$\sim$2.1 MeV. This first estimate of the resonant rate shows that the first excited state ($E_x$=0.373 MeV) dominates the resonant
contribution below 0.2 GK, the second excited state ($E_x$=0.593 MeV) dominates around 0.2--1.7 GK, and the high-lying 2.067 MeV state
(with much shorter life-time $\tau$=30 fs) dominates at even higher temperature. It shows that the contribution owing to those
high-lying states above 2.067 MeV is negligible at temperatures of interest in XRBs.

We then improved upon this first estimate of the resonant rate.
The simplest model for calculating the isobaric-multiplet-mass-equation (IMME) is the $0f_{7/2}$ shell model used in~\cite{bib:bro79}
where the displacement energies in the mass region A=41-55 were used to deduce the effective isovector and isotensor two-body matrix
elements. The root-mean-square difference between experiment and theory for 60 $\Delta Z$=1 displacement energies was 12 keV. With
this model the $\Delta Z$=3 displacement energy difference between $^{43}$Ca and $^{43}$V (7/2$^-$) state is predicted to be
22.854(36) MeV compared to the new experimental value of 22.857(43) MeV. The agreement is impressive.
In the framework of an OXBASH~\cite{bib:oxbash} shell model, the resonant parameters of the three states discussed above have been
recalculated and summarized in Table~\ref{table1}. These calculations are discussed in detail in Appendix A.

\begin{table*}
\caption{\label{table1} Parameters for the present $^{42}$Ti($p$,$\gamma$)$^{43}$V resonant rate calculation. The uncertainties quoted for
strengths ($\omega\gamma$) arise from the energy dependence of the widths $\Gamma_\gamma$ and $\Gamma_p$, as well as the assumed
uncertainties of spectroscopic factors ($C^2S_p$) (a factor of 2).}
\begin{ruledtabular}
\begin{center}
\begin{tabular}{|c|c|c|c|c|c|c|c|c|}
$E_x$($^{43}V$) (MeV) & $E_r$ (MeV)\footnotemark[1] & $\tau$ (ps) & $J^{\pi}$	& $\ell$ & $C^2S_p$ & $\Gamma_\gamma$ (eV) & $\Gamma_p$ (eV) & $\omega\gamma$ (eV) \\
\hline
0.436(0.050)\footnotemark[2] & 0.353(0.066) & 22(2)	     & 5/2$^-$ & 3 & 0.15\footnotemark[3]   & 3.04$\times$10$^{-5}$ & 5.10$\times$10$^{-9}$ & 1.5$\times$10$^{-8}$ \\
                             &              &     	     &         &   &                        &                       &                       & lower: 1.4$\times$10$^{-10}$ \\
                             &              &     	     &         &   &                        &                       &                       & upper: 6.3$\times$10$^{-7}$ \\
\hline
0.537(0.050)\footnotemark[2] & 0.454(0.066) & 117(6)	 & 3/2$^-$ & 1 & 0.046\footnotemark[4]  & 3.42$\times$10$^{-6}$ & 6.27$\times$10$^{-5}$ & 6.5$\times$10$^{-6}$ \\
                             &              &     	     &         &   &                        &                       &                       & lower: 2.2$\times$10$^{-6}$ \\
                             &              &     	     &         &   &                        &                       &                       & upper: 1.1$\times$10$^{-5}$ \\
\hline
2.067(0.100)\footnotemark[2] & 1.984(0.109) & 0.03(0.01) & 7/2$^-$ & 3 & 0.0003\footnotemark[5] & 2.19$\times$10$^{-2}$ & 3.45$\times$10$^{-2}$ & 5.4$\times$10$^{-2}$ \\
                             &              &     	     &         &   &                        &                       &                       & lower: 2.2$\times$10$^{-2}$ \\
                             &              &     	     &         &   &                        &                       &                       & upper: 1.1$\times$10$^{-1}$ \\
\end{tabular}
\end{center}
\footnotemark[1] Resonance energies calculated using $E_r$=$E_x$($^{43}V$)-$S_p^{\mathrm{exp}}$, where $S_p^{\mathrm{exp}}$=83$\pm$43 keV.
\footnotemark[2] Estimated theoretical uncertainties in the parenthesis.
\footnotemark[3] Value from the previous ($p$,$d$)~\cite{bib:sam68a} and ($d$,$t$)~\cite{bib:do76} experiments.
\footnotemark[4] Averaged value from the ($d$,$p$) experiments~\cite{bib:do66b,bib:br74b}.
\footnotemark[5] Value calculated by the OXBASH code with same model-space and interactions as in Ref.\cite{bib:her95}.
\end{ruledtabular}
\end{table*}

\subsection{Direct capture rate}
The nonresonant direct capture (DC) rate can be estimated using methods presented in Refs.~\cite{bib:rol88,bib:her95},
\begin{widetext}
\begin{eqnarray}
N_A\langle \sigma v \rangle = N_A\left (\frac{8}{\pi A}\right)^{1/2}\frac{1}{(kT)^{3/2}}\int^{\infty}_{0}S_\mathrm{dc}(E) \mathrm{exp} \left[-\frac{E}{kT}-\frac{b}{E^{1/2}} \right]dE
\label{eq5}
\end{eqnarray}
\end{widetext}
If $S_\mathrm{dc}(E)$ factor is nearly a constant over the Gamow window, the nonresonant reaction rate can be approximated
in a form of~\cite{bib:rol88,bib:her95}
\begin{widetext}
\begin{eqnarray}
N_A\langle \sigma v \rangle_\mathrm{dc}=7.83 \times 10^{9} \left( \frac{Z}{A} \right)^{1/3}T_9^{-2/3}S_\mathrm{dc}(E_0) \mathrm{[MeV~b]}\times \mathrm{exp} \left[ -4.249 \left (\frac{Z^2 A}{T_9} \right)^{1/3} \right] [\mathrm{cm^3s^{-1}mol^{-1}}].
\label{eq6}
\end{eqnarray}
\end{widetext}
The critical parameter is $S_\mathrm{dc}(E_0)$, the astrophysical $S$-factor at the Gamow energy $E_0$. Herndl \textit{et al.} listed
an effective $S_\mathrm{dc}(E_0)$ factor of 4.91$\times$10$^{-20}$ [MeV b] in their Table XIII. We have recalculated this factor and
found that the above number is actually 4.91$\times$10$^{-2}$ [MeV b].

In this work, the $^{42}$Ti($p$,$\gamma$)$^{43}$V reaction rate from direct capture into ground state of $^{43}$V has been
calculated with a RADCAP code~\cite{bib:ber03,bib:hua10} by using a Woods-Saxon nuclear potential (central + spin orbit) and a
Coulomb potential of a uniform charge distribution. The nuclear potential parameters were determined by matching the bound-state
energy ($E_b$=83 keV). A spectroscopic factor of $C^2S$=0.75~\cite{bib:her95}, which agrees with the ($d$,$p$) experimental values of
0.68~\cite{bib:do66b} and 0.55~\cite{bib:br74b}, was adopted in the present calculations.
The DC rate contributes to the total rate only by 10--20\% in the temperature region of 2--3 GK, and dominates the rate below 0.07 GK.
The RADCAP calculations are described in detail in Appendix B.

\subsection{Total reaction rate}
\begin{figure}[t]
\begin{center}
\includegraphics[width=8.8cm]{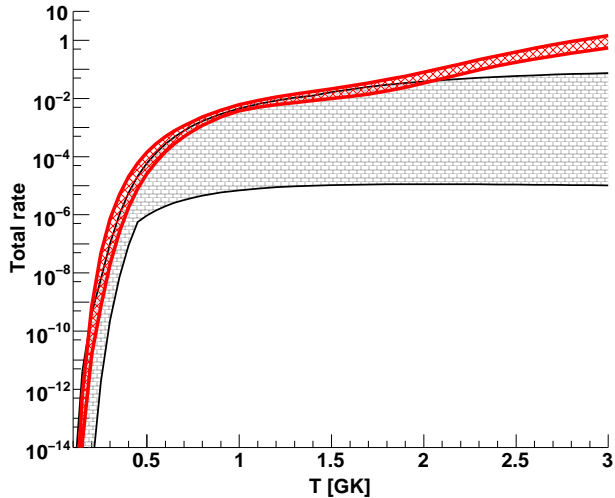}
\vspace{-6mm}
\caption{\label{fig1} (Color online) Total reaction rate calculated for the $^{42}$Ti($p$,$\gamma$)$^{43}$V reaction (in units of cm$^3$ mole$^{-1}$ s$^{-1}$).
The upper and lower limits of the present rate (with $S_p$=83$\pm$43 keV~\cite{bib:yan13}) are shown by the (red) thicker lines, and those of the
Herndl \textit{et al.} rate (with $S_p$=90$\pm$200 keV~\cite{bib:her95}) are shown by the black thin lines. See text for details.}
\end{center}
\end{figure}

The total reaction rate of $^{42}$Ti($p$,$\gamma$)$^{43}$V has been calculated by simply summing up the resonant and DC contributions.
Our new rate is tabulated in Table~\ref{table2} and plotted in Fig.~\ref{fig1}. The uncertainty in the present rate arises from
uncertainties in our adopted $E_r$ (which also lead to uncertainties in the strengths since the values of $\Gamma_p$ and $\Gamma_\gamma$
have been scaled using the values of $E_r$ - see Appendix A) and the uncertainty in the DC contribution ($\approx$40\% - see Appendix B).
In addition, we have assumed a factor of two uncertainty in the adopted spectroscopic factors. The uncertainty of the total rate is dominated
by the uncertainty of the $S_p$ value due to the exponential dependence of the rate on $S_p$.
The rate based upon calculations in Herndl {\it et al.}, where only two states (at $E_x$=0.36, 0.55 MeV) were assumed, is also shown in
Fig.~\ref{fig1} for comparison.
Because the uncertainty of the DC contribution was not determined in Herndl {\it et al.}, the uncertainty of Herndl {\it et al.} rate
shown originates only from those of the calculated resonant rates ({\it i.e.}, the error of $S_p$ propagating into the strengths).
It shows our new rate calculated with the precise experimental $S_p$ value has much smaller uncertainties than the
previous ones. This clearly demonstrates the importance of precise mass measurements.

\begin{table}
\caption{\label{table2} Reaction rates calculated for $^{42}$Ti($p$,$\gamma$)$^{43}$V. All the rates are in units of cm$^3$ mole$^{-1}$ s$^{-1}$.}
\begin{ruledtabular}
\begin{center}
\begin{tabular}{|c|c|c|c|}
T [GK] & DC & Resonant & Total	\\
\hline
0.01 &	3.69$\times$10$^{-57}$ &	2.96$\times$10$^{-178}$ &	3.69$\times$10$^{-57}$ \\
0.02 &	1.36$\times$10$^{-43}$ &	9.44$\times$10$^{-90}$  &	1.36$\times$10$^{-43}$ \\
0.03 &	5.18$\times$10$^{-37}$ &	2.31$\times$10$^{-60}$  &	5.18$\times$10$^{-37}$ \\
0.04 &	7.32$\times$10$^{-33}$ &	1.00$\times$10$^{-45}$  &	7.32$\times$10$^{-33}$ \\
0.05 &	6.56$\times$10$^{-30}$ &	5.65$\times$10$^{-37}$  &	6.56$\times$10$^{-30}$ \\
0.06 &	1.17$\times$10$^{-27}$ &	3.66$\times$10$^{-31}$  &	1.17$\times$10$^{-27}$ \\
0.08 &	2.25$\times$10$^{-24}$ &	6.15$\times$10$^{-24}$  &	8.40$\times$10$^{-24}$ \\
0.09 &	4.04$\times$10$^{-23}$ &	1.53$\times$10$^{-21}$  &	1.57$\times$10$^{-21}$ \\
0.10 &	4.87$\times$10$^{-22}$ &	1.24$\times$10$^{-19}$  &	1.24$\times$10$^{-19}$ \\
0.20 &	8.20$\times$10$^{-16}$ &	7.58$\times$10$^{-11}$  &	7.58$\times$10$^{-11}$ \\
0.30 &	8.64$\times$10$^{-13}$ &	1.64$\times$10$^{-7}$   &	1.64$\times$10$^{-7}$ \\
0.40 &	7.01$\times$10$^{-11}$ &	8.05$\times$10$^{-6}$   &	8.05$\times$10$^{-6}$ \\
0.50 &	1.60$\times$10$^{-9}$  &	7.87$\times$10$^{-5}$   &	7.87$\times$10$^{-5}$ \\
0.60 &	1.75$\times$10$^{-8}$  &	3.44$\times$10$^{-4}$   &	3.44$\times$10$^{-4}$ \\
0.70 &	1.18$\times$10$^{-7}$  &	9.54$\times$10$^{-4}$   &	9.54$\times$10$^{-4}$ \\
0.80 &	5.75$\times$10$^{-7}$  &	2.00$\times$10$^{-3}$   &	2.00$\times$10$^{-3}$ \\
0.90 &	2.19$\times$10$^{-6}$  &	3.47$\times$10$^{-3}$   &	3.47$\times$10$^{-3}$ \\
1.00 &	2.19$\times$10$^{-6}$  &	5.32$\times$10$^{-3}$   &	5.32$\times$10$^{-3}$ \\
1.10 &	1.90$\times$10$^{-5}$  &	7.44$\times$10$^{-3}$   &	7.46$\times$10$^{-3}$ \\
1.20 &	4.66$\times$10$^{-5}$  &	9.75$\times$10$^{-3}$   &	9.79$\times$10$^{-3}$ \\
1.30 &	1.04$\times$10$^{-4}$  &	1.22$\times$10$^{-2}$   &	1.23$\times$10$^{-2}$ \\
1.40 &	2.15$\times$10$^{-4}$  &	1.48$\times$10$^{-2}$   &	1.50$\times$10$^{-2}$ \\
1.50 &	4.15$\times$10$^{-4}$  &	1.77$\times$10$^{-2}$   &	1.81$\times$10$^{-2}$ \\
2.00 &	5.61$\times$10$^{-3}$  &	5.60$\times$10$^{-2}$   &	6.16$\times$10$^{-2}$ \\
2.50 &	3.65$\times$10$^{-2}$  &	2.46$\times$10$^{-1}$   &	2.82$\times$10$^{-1}$ \\
3.00 &	1.55$\times$10$^{-1}$  &	7.90$\times$10$^{-1}$   &	9.45$\times$10$^{-1}$ \\
\end{tabular}
\end{center}
\end{ruledtabular}
\end{table}

\begin{figure}[h]
\begin{center}
\includegraphics[width=6.4cm]{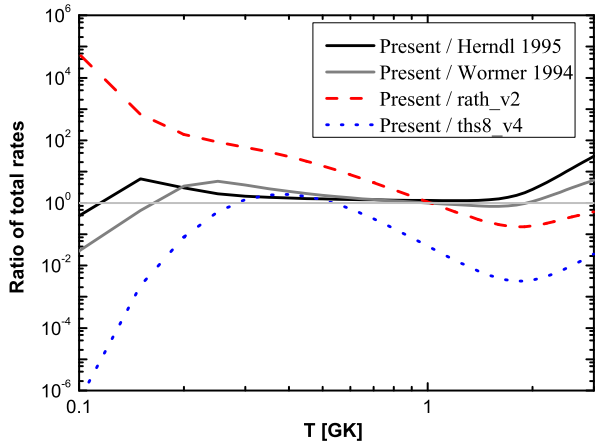}
\vspace{-2mm}
\caption{\label{fig2} (Color online) Ratios between the present rate  (see Table~\ref{table2}) and other available ones (Herndl 1995~\cite{bib:her95}, Wormer 1994~\cite{bib:wor94},
rath\_v2~\cite{bib:jina} and ths8\_v4~\cite{bib:jina}).}
\end{center}
\end{figure}

Figure~\ref{fig2} compares five different rates for the $^{42}$Ti($p$,$\gamma$)$^{43}$V reaction:
(a) present rate (Fig.~\ref{fig1}(a)); (b) the rate from Herndl {\it et al.}~\cite{bib:her95}; (c) the rate from Wormer {\it et al.}~\cite{bib:wor94};
(d) the statistical model rate ths8\_v4 available in the JINA REACLIB (with $S_p$=-0.0189 MeV~\cite{bib:cyb10}); (e) the statistical model
rate rath\_v2 in the REACLIB~\cite{bib:jina} (with $S_p$=-0.411 MeV based on the FRDM mass model~\cite{bib:frdm}).
Because of the rather similar $S_p$ value used, our new rate does not deviate significantly from those of Herndl {\it et al.} and Wormer {\it et al.} in
the temperature region of interest in XRBs. Our new rate, however, is very well constrained with the precise mass measurement as shown in
Fig.~\ref{fig1}. The statistical-model calculations deviate from our new rate considerably over the entire temperature region of interest.
This demonstrates again that the statistical-model is not ideally applicable for this reaction mainly owing to the low density of
low-lying excited states in $^{43}$V.

\section{Astrophysical implications}
The impact of our new $^{42}$Ti($p$,$\gamma$)$^{43}$V rate was examined in the framework of one-zone XRB models. Using the
representative K04 thermodynamic history ($T_\mathrm{peak}$=1.4 GK~\cite{bib:par08}), we performed a series of postprocessing
calculations to explore the role of different $^{42}$Ti($p$,$\gamma$)$^{43}$V rates and $S_p$ values on the nuclear energy
generation rate ($E_\mathrm{nuc}$) and XRB yields. Rates of all other reactions in the network were left unchanged during these
calculations. To be clear, in the discussion below we will refer explicitly to $^{42}$Ti($p$,$\gamma$)$^{43}$V forward rates
({\it e.g.}, as shown in Fig.~\ref{fig1}) and to the $S_p$ value used to determine the corresponding reverse rates through
the principle of detailed balance (see, {\it e.g.},~\cite{bib:par09}).

No significant differences in the respective nuclear energy generation rates were found by comparing XRB calculations
with the (a) present forward rate ($S_p$=83 keV for the reverse rate); (b) Herndl \textit{et al.} forward rate ($S_p$=88 keV);
(c) Wormer \textit{et al.} forward rate ($S_p$=88 keV); and (d) ths8\_v4 forward rate
($S_p$=-19 keV). $E_\mathrm{nuc}$ determined using the rath\_v2 forward rate ((e), $S_p$=-411 keV), however, was up to 10\%
lower than that from the above cases (a--d) during the burst. This (minor) difference is attributed to the very
different $S_p$ value used in the rath\_v2 reverse rate: the $E_\mathrm{nuc}$ from an additional XRB calculation
(f) performed with a reverse rate recalculated using the rath\_v2 forward rate and $S_p$=83 keV agreed well with the
$E_\mathrm{nuc}$ from cases (a--d) above. This is because an equilibrium between the forward
$^{42}$Ti($p$,$\gamma$)$^{43}$V and reverse $^{43}$V($\gamma$,$p$)$^{42}$Ti processes is quickly established owing to the
relatively small $S_p$(=83 keV) of $^{43}$V relative to XRB temperatures ({\it e.g.}, at 1 GK, $kT$$\approx$100 keV).
As a result, the actual rate of the $^{42}$Ti($p$,$\gamma$)$^{43}$V reaction is only of secondary importance;
instead, it is the reaction $Q$-value (or $S_p$ value) that characterizes the equilibrium abundances of $^{42}$Ti and $^{43}$V
and the energy release through subsequent reactions on these species.

\begin{figure}[h]
\begin{center}
\includegraphics[width=8.5cm]{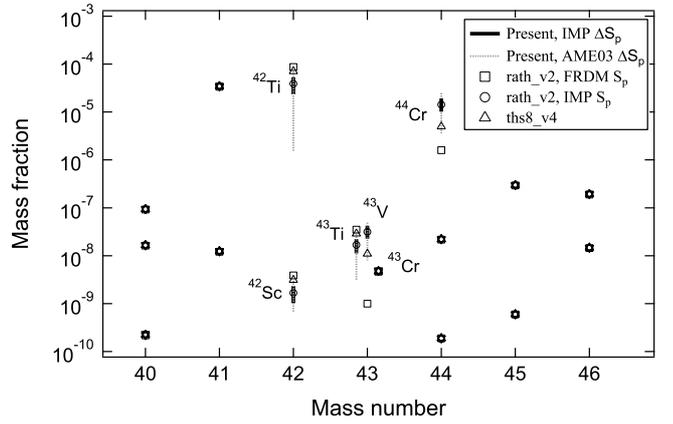}
\vspace{-1mm}
\caption{\label{fig3} Abundances following one-zone XRB calculations using the K04 thermodynamic history~\cite{bib:par08}.
Abundance variations determined using the present $^{42}$Ti($p$,$\gamma$)$^{43}$V forward rate with reverse rates calculated
using $\Delta S_p$=43 keV (IMP~\cite{bib:yan13}, solid black line), and $\Delta S_p$=233 keV (AME03~\cite{bib:aud03}, dotted grey line)
are indicated. As well, abundances determined using the rath\_v2 forward rate~\cite{bib:jina} along with reverse
rates calculated with $S_p$=-411 keV (FRDM~\cite{bib:frdm}, open squares) and $S_p$=83 keV (IMP~\cite{bib:yan13}, open circles) are shown.
Abundances determined with the ths8\_v4 rate~\cite{bib:jina} ($S_p$=-19 keV~\cite{bib:cyb10}, open triangles) are also shown.}
\end{center}
\end{figure}

\begin{table*}
\caption{\label{table3} Mass fractions following one-zone XRB calculations using the K04 thermodynamic history~\cite{bib:par08}.
These values are plotted in Fig.~\ref{fig3}. The first two columns give ranges of mass fractions as determined using the one
sigma uncertainties for $S_p$($^{43}$V) from the recent measurement and other theoretical estimates.}
\begin{ruledtabular}
\begin{center}
\begin{tabular}{|c|c|c|c|c|c|}
\multicolumn{1}{|c|}{} &\multicolumn{5}{c|}{Mass fraction} \\
\hline
Species   & IMP $\Delta S_p$~\cite{bib:yan13} &    AME03 $\Delta S_p$~\cite{bib:aud03} & rath\_v2 (FRDM $S_p$~\cite{bib:frdm}) &  rath\_v2 (IMP $S_p$)& ths8\_v4~\cite{bib:cyb10} \\
\hline
$^{42}$Ti & (2.3--5.3)$\times$10$^{-5}$	& (1.6--75)$\times$10$^{-6}$  & 8.7$\times$10$^{-5}$  & 3.9$\times$10$^{-5}$ & 7.1$\times$10$^{-5}$ \\
$^{42}$Sc & (1.0--2.3)$\times$10$^{-9}$	& (6.9--33)$\times$10$^{-10}$ &	3.9$\times$10$^{-9}$  &	1.7$\times$10$^{-9}$ & 3.1$\times$10$^{-9}$ \\
$^{43}$Ti & (1.1--2.2)$\times$10$^{-8}$	& (3.2--31)$\times$10$^{-9}$  & 3.5$\times$10$^{-8}$  &	1.7$\times$10$^{-8}$ & 2.9$\times$10$^{-8}$ \\
$^{43}$V  & (2.3--4.2)$\times$10$^{-8}$	& (8.0--55)$\times$10$^{-9}$  &	1.0$\times$10$^{-9}$  &	3.2$\times$10$^{-8}$ & 1.1$\times$10$^{-8}$ \\
$^{44}$Cr & (1.0--1.9)$\times$10$^{-5}$	& (3.6--25)$\times$10$^{-6}$  &	1.6$\times$10$^{-6}$  &	1.4$\times$10$^{-5}$ & 5.0$\times$10$^{-6}$ \\
\end{tabular}
\end{center}
\end{ruledtabular}
\end{table*}

The effects on XRB yields by using different $^{42}$Ti($p$,$\gamma$)$^{43}$V forward rates and $S_p$ values have been investigated.
Fig.~\ref{fig3} shows representative yields in this mass range for the different cases discussed above, as determined immediately
following the respective XRB calculations. No significant differences in yields were
observed for cases (a--c,f) above. The two cases (d,e) with reverse rates determined using negative $S_p$ values gave
somewhat different yields for species with A=42--44. For example, the negative $S_p$ values produce relatively more $^{42}$Ti
but less $^{43}$V.

The dominant role of the $S_p$ value used in the reverse rate in determining the yields is clearly seen in Fig.~\ref{fig3}
from the comparison of cases (a) (labeled as ``Present, IMP $\Delta S_p$"), (d) (labeled as ``ths8\_v4"), (e) (labeled as
``rath\_v2, FRDM $S_p$"), and (f) (labeled as ``rath\_v2, IMP $S_p$"). It shows the yields calculated with the new
experimental $S_p$ value (for the reverse rate) significantly differ from those yields with other theoretical $S_p$ values.
In addition, to demonstrate the impact of the uncertainty in $S_p$, we performed additional XRB calculations using the present
forward rate, along with reverse rates that reflect the one sigma uncertainties in $S_p$ from AME03 ($\Delta S_p$=233 keV) and the IMP
mass measurement ($\Delta S_p$=43 keV). As shown in Fig.~\ref{fig3}, the reduced uncertainty in $S_p$ directly influences the
possible ranges of mass fractions for the affected species. Indeed, the uncertainty from the IMP mass measurement leads to
variations, by less than a factor of three, in the yields of the most produced isotopes in this mass region, such as $^{42,43}$Ti, $^{42}$Sc,
$^{43}$V and $^{43,44}$Cr.

\section{Summary}
The thermonuclear rate of the $^{42}$Ti($p$,$\gamma$)$^{43}$V reaction has been recalculated using the recent precise proton
separation energy of $S_p$=83$\pm$43 keV measured at the HIRFL-CSR facility in Lanzhou, China. We have also used new, updated
calculations of the direct capture and resonant contributions to the rate. Our new rate deviates significantly from other rates found
in the literature. We confirm that statistical model calculations are not ideally applicable for this reaction primarily because of the
low density of low-lying excited states in $^{43}$V. We recommend that out new rate be incorporated in future astrophysical
network calculations.

The astrophysical impact of our new rate has been investigated through one-zone postprocessing Type I x-ray burst calculations.
Even when using dramatically different rates, we find no significant changes to the calculated nuclear energy generation rate during a
representative burst. This is because equilibrium between the forward $^{42}$Ti($p$,$\gamma$)$^{43}$V and reverse $^{43}$V($\gamma$,$p$)$^{42}$Ti
processes rapidly develops at XRB temperatures. As such it is the reaction $Q$-value (or $S_p$) that mainly characterizes
the equilibrium abundances of $^{42}$Ti and $^{43}$V. In this respect, the present $^{42}$Ti($p$,$\gamma$)$^{43}$V rate and $S_p$($^{43}$V)
value are sufficiently well known to determine the nuclear energy generation rate within the framework of the adopted XRB model.
In addition, we find that the new experimental value of $S_p$ affects significantly the yields of a limited number of species with
A=42--44, such as $^{42,43}$Ti, $^{42}$Sc, $^{43}$V and $^{43,44}$Cr. The precision in $S_p$ achieved from the IMP mass measurement
restricts the variation of these yields to better than a factor of three. It demonstrates clearly the importance
of precise mass measurements for those key nuclei (especially those waiting-point nuclei) along the rp-process occurring
in x-ray bursts.

\begin{center}
\textbf{Acknowledgments}
\end{center}
This work was financially supported by the National Natural Science Foundation of China (Nos. 11135005, U1232208), the Major State
Basic Research Development Program of China (2013CB834406, 2013CB834401).
AP was supported by the Spanish MICINN (Nos. AYA2010-15685, EUI2009-04167), by the E.U. FEDER funds as well as by the ESF
EUROCORES Program EuroGENESIS, BAB was supported by the NSF grant (PHY-1068217), and TR was supported by the Swiss NSF, EuroGENESIS
and the ENSAR/THEXO collaboration within the 7th Framework Programme of the EU.

\appendix
\section{Calculation of resonant parameters}
Below we summarize our calculations of the resonant parameters of the three states in $^{43}$V around 0.373 MeV (Resonance 1),
0.593 MeV (Resonance 2) and 2.067 MeV (Resonance 3).
\subsection{Resonance 1}
In the $0f_{7/2}$ model, the energy of the first excited 5/2$^-$ state is predicted to be 63 keV higher in $^{43}$V compared to $^{43}$Ca.
Thus, with the experimental energy of 0.370 keV for $^{43}$Ca we obtain a predicted excitation energy of 0.436 MeV ($E_r$=0.353 MeV)
for $^{43}$V.

The 5/2$^-$ to 7/2$^-$ transition in $^{43}$Ca has an experimental $B(M1)$ value of 0.023(2) $\mu_N^2$. In the $pf$ model space with
the FPD6 interaction~\cite{bib:fpd6} the $B(M1)$ values are predicted to be 0.018, 0.025 $\mu_N^2$ for the excited states in
$^{43}$Ca and $^{43}$V, respectively. Therefore, a value of $B(M1)$=0.032 $\mu_N^2$ is derived for the predicted 0.436 MeV state, and
$\Gamma_\gamma$ is thus calculated to be about 3.04$\times$10$^{-5}$ eV. This implies that the resonance strength for
this state is determined by the much smaller $\Gamma_p$.

Wormer \textit{et al.} estimated a resonant strength value of $\omega\gamma=$1.0$\times$10$^{-10}$ eV for the
$E_r$=0.28 MeV state.
A spectroscopic factor of 0.014 would reproduce their proton width of $\Gamma_p$$\approx$3.3$\times$10$^{-11}$ eV based on Eq.~\ref{eq3}.
Later, Herndl {\it et al.} estimated a value of $\Gamma_p$=1.1$\times$10$^{-12}$ eV with a spectroscopic factor of 0.008, by the equation of
\begin{eqnarray}
\Gamma_{p}=C^2S_p \times \Gamma_\mathrm{sp},
\label{eq4}
\end{eqnarray}
where the single-particle width $\Gamma_\mathrm{sp}$ was calculated from the scattering phase shifts in a Woods-Saxon
potential~\cite{bib:bro93,bib:cha93} whose depth was determined by matching the resonant energy. Based on the method introduced in
Ref.~\cite{bib:ili97}, we have recalculated this proton width using using Equ.~\ref{eq4} with potential parameters of $E_r$=0.27 MeV,
$r_0$=1.17 fm, $a$=0.69 fm and $r_c$=1.28 fm. The justification of the choice of the parameters can be found in ~\cite{bib:ili97}.
We have obtained a single-particle width of $\Gamma_\mathrm{sp}$=1.76$\times$10$^{-10}$ eV, which roughly agrees with the value of
1.38$\times$10$^{-10}$ eV calculated by Herndl \textit{et al.}

Neutron spectroscopic factor measurements imply values of $S_n$$\approx$0.15~\cite{bib:sam68a,bib:do76}, in disagreement with
the values assumed by Wormer {\it et al.} and Herndl {\it et al.} We assumed $S_p$=$S_n$ in the following proton width calculations.
With Equ.~\ref{eq4}, a width of $\Gamma_p$=5.10$\times$10$^{-9}$ eV was obtained with the above parameters of $r_0$=1.17 fm, $a$=0.69 fm
and $r_c$=1.28 fm (parameter Set~3 in Table~\ref{table4}). The proton width calculated by Eq.~\ref{eq3} is always larger than that
by Eq.~\ref{eq4}, because the former equation does not take the dimensionless single-particle reduced width
$\theta_\mathrm{sp}^2$~\cite{bib:ili97} into account. $\theta_\mathrm{sp}^2$ is usually assumed to be unity, although this is not
appropriate for many cases~\cite{bib:ili97}. Here, $\theta_\mathrm{sp}^2$ is calculated to be 0.24.

\subsection{Resonance 2}
A description of the 3/2$^-$ excited state requires the full $pf$ shell-model basis. With the empirically determined
isospin-nonconserving interactions for the $pf$ shell~\cite{bib:orm89}, the second excited 3/2$^-$ state in $^{43}$V is
estimated to be located at $E_x$=0.537 MeV ($E_r$=0.454 MeV).

A spectroscopic factor of $C^2S$=0.046 averaged from the ($d$,$p$) experiments~\cite{bib:do66b,bib:br74b} was used for this state, as
adopted in Ref.~\cite{bib:wor94}. The proton
width is calculated to be 6.27$\times$10$^{-5}$ eV ($\theta_\mathrm{sp}^2$=0.56) with the same parameter Set~3 (Table~\ref{table4}).
Since it is difficult to make a reliable life-time calculation for this state, we estimated this $\Gamma_\gamma$ based on the mirror
life-time. In the mirror $^{43}$Ca, this state decays either to the ground state ($J^{\pi}$=7/2$^-$) or to the first excited state
($J^{\pi}$=5/2$^-$) with branching ratios~\cite{bib:end90} of 70.2\% and 29.8\%, respectively. The ground-state transition is
a pure $E2$, whose width can be estimated by the relation of
$\Gamma_\gamma(E2)$=$S$$\times$$\Gamma^W_\gamma(E2)$$\times$$BR$~\cite{bib:end79}. Here, $S$ is the strength of the transition in
Wiesskopf units, and $BR$ is the branching ratio (70.2\%). The Weisskopf-unit gamma width (in eV) for an $E2$ transition is
$\Gamma^W_\gamma(E2)$=4.9$\times$10$^{-8}$A$^{4/3}$$E_\gamma^5$~\cite{bib:end79,bib:wil60} with A=43. This results in a
ground-state-transition width of $\Gamma_\gamma(E2)$$\approx$1.7$\times$10$^{-6}$ eV with $S$=7.2~\cite{bib:end79}. The
first-excited-state transition is a mixture of $M1$ and $E2$, where the dominant $M1$ width can be calculated by the relation of
$\Gamma_\gamma(M1)$=$S$$\times$$\Gamma^W_\gamma(M1)$$\times$$BR$. Here, a value of $S$=7.6$\times$10$^{-3}$~\cite{bib:end79}
was adopted in the calculation. The Weisskopf-unit gamma width (in eV) for an $M1$ transition is
$\Gamma^W_\gamma(M1)$=2.1$\times$10$^{-2}$$E_\gamma^3$~\cite{bib:end79,bib:wil60}. $\Gamma_\gamma(M1)$ is estimated to be about
4.9$\times$10$^{-7}$ eV with a branching ratio of 29.8\%. Therefore, only the ground-state-transition dominates the actual total
$\Gamma_\gamma$ width, and the energy dependence of $\Gamma_\gamma$ can be accounted for by using the scale factor $E_\gamma^5$.
For the 0.593-MeV state in $^{43}$Ca, $\Gamma_\gamma$ is about 5.62$\times$10$^{-6}$ eV (as estimated from the lifetime of 117 ps)
In this work, we have adopted a value of $\Gamma_\gamma$=3.42$\times$10$^{-6}$ eV for the 0.537-MeV state in $^{43}$V by correcting
for the energy difference between $^{43}$V ($E_x$=0.537 MeV) and $^{43}$Ca ($E_x$=0.593 MeV).

\subsection{Resonance 3}
The higher-lying 2.067-MeV 7/2$^-$ state in $^{43}$Ca is not described in the $pf$ model space, and requires nucleons to be excited
from the $sd$ shell for its description. We do not have a good model for its displacement energy and simply use the same value for its
excitation energy in $^{43}$V with an estimated error of 100 keV.

The $\Gamma_\gamma$ for this state was calculated to be 2.19$\times$10$^{-2}$ eV with a mirror life-time of
$\tau$=0.03 ps. In the mirror $^{43}$Ca, this state mainly decays to the ground state ($J^{\pi}$=7/2$^-$) and to the first excited state
($J^{\pi}$=5/2$^-$) with branching ratios~\cite{bib:end90} of 78\% and 22\%, respectively; both $\gamma$ transitions have $M1$($E2$)
characters. By using the same strength $S$ value for the above 0.537 MeV state with respect to $E2$ and $M1$ transitions,
$\gamma$ widths of the ground-state and first-excited transitions were calculated. It is found that the ground-state $E2$ transition
dominates the total $\Gamma_\gamma$ for this state.
Therefore, the factor $E_\gamma^5$ was again used to account for the energy dependence of $\Gamma_\gamma$.
The proton width $\Gamma_p$ was calculated to be 3.45$\times$10$^{-2}$ eV with parameter Set~3 (Table~\ref{table4}).
We have used a spectroscopic factor of 0.0003 as determined with the OXBASH code (using the same model-space and interactions as in
Ref.~\cite{bib:her95}). This factor may be larger in nature, and should be determined experimentally.

\section{Calculation of direct capture rate}

\begin{table*}
\caption{\label{table4} Potential parameter lists used in the $S_\mathrm{dc}$ factor calculations.}
\begin{ruledtabular}
\begin{center}
\begin{tabular}{|c|c|c|c|}
Parameters & Set 1~\cite{bib:hua10} & Set 2~\cite{bib:per63}\footnotemark[2] & Set 3~\cite{bib:ili97}\footnotemark[2]	\\
\hline
$R_0$=$R_{so}$ (fm)\footnotemark[1] & 1.25$\times$(1+42)$^{\frac{1}{3}}$ [$r_0$=$r_{so}$=1.26] & 1.25$\times$42$^{\frac{1}{3}}$ [$r_0$=$r_{so}$=1.25] & 1.25$\times$42$^{\frac{1}{3}}$-0.23 [=1.17$\times$42$^{\frac{1}{3}}$, $r_0$=$r_{so}$=1.17] \\
$R_c$ (fm)\footnotemark[1]          & 1.25$\times$(1+42)$^{\frac{1}{3}}$ [$r_c$=1.26]          & 1.25$\times$42$^{\frac{1}{3}}$ [$r_c$=1.25]          & 1.24$\times$42$^{\frac{1}{3}}$+0.12 [=1.28$\times$42$^{\frac{1}{3}}$, $r_c$=1.28]          \\
$a_0$=$a_{so}$ (fm) & 0.65                               & 0.65                           & 0.69                                                                  \\
$V_{so}$ (MeV)      & -10.0                              &  -10.0                         &  -10.0                                                                \\
\hline
$V_{0}$ (MeV)\footnotemark[3] & -100.48                  &  -101.73                       &  -111.22                                                              \\
$S_\mathrm{dc}(0)$ (MeV b)    & 3.98$\times$10$^{-2}$    &  3.84$\times$10$^{-2}$         &  3.48$\times$10$^{-2}$                                                \\
\end{tabular}
\end{center}
\footnotemark[1] $r_0$, $r_{so}$ and $r_c$ are commonly defined as $r$=$R$/$A^{\frac{1}{3}}$ for comparison.
\footnotemark[2] The choice of parameters can be found in Ref.~\cite{bib:ili97}.
\footnotemark[3] $V_0$ is varied to match the bound-state energy $E_b$=83 keV.
\end{ruledtabular}
\end{table*}

\begin{figure}
\includegraphics[width=8.8cm]{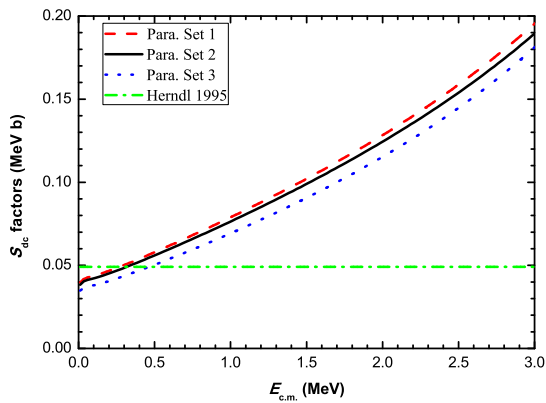}
\vspace{-7mm}
\caption{\label{fig4} (Color online) Direct-capture $S_\mathrm{dc}$ factors calculated with three parameter sets listed in Table~\ref{table4}.
A previous constant value of $S_\mathrm{dc}(E_0)$=4.91$\times$10$^{-2}$ [MeV b] (Herndl 1995~\cite{bib:her95}) is shown for comparison.}
\end{figure}

The astrophysical $S$ factor of the direct-capture $^{42}$Ti($p$,$\gamma$)$^{43}$V reaction has been calculated by the RADCAP code.
The calculated $S_\mathrm{dc}$ factors are shown in Fig.~\ref{fig4} with three parameter sets listed in Table~\ref{table4}.
With a spin-orbit potential of $V_{so}$=-10 MeV, the $S_\mathrm{dc}(E)$ factors calculated using three sets of parameters
(Table~\ref{table4}) vary by no more than 15\% over the energy range of 0--3 MeV.
This energy range covers the Gamow window for a temperature up to 3 GK. The above changes can not be regarded as substantial.
Since Huang \textit{et al.}~\cite{bib:hua10} reproduced successfully the $S$ factors for a series of radiative capture reactions, we have
adopted their potential parameters (Set 1 in Table~\ref{table4}) in the final DC rate calculation. The present $S_\mathrm{dc}$ factors can be well
parameterized in a Taylor-series form~\cite{bib:rol88} of $S_\mathrm{dc}(E)$=$\sum\limits_{k=1}^k$$\frac{S^{(k)}(0)}{k!}$$E^k$, where $S$
factors are in units of [MeV b] and $E$ in MeV. The fitted parameters are $S(0)$=3.97$\times$10$^{-2}$ [MeV b] for the $S$ factor at zero energy,
and the derivatives with respect to energy are $S^{(1)}(0)$=3.37$\times$10$^{-2}$, $S^{(2)}(0)$=1.31$\times$10$^{-2}$,
$S^{(3)}(0)$=9.72$\times$10$^{-3}$ and $S^{(4)}(0)$=1.18$\times$10$^{-2}$, respectively.

\begin{figure*}
\begin{center}
\includegraphics[width=12cm]{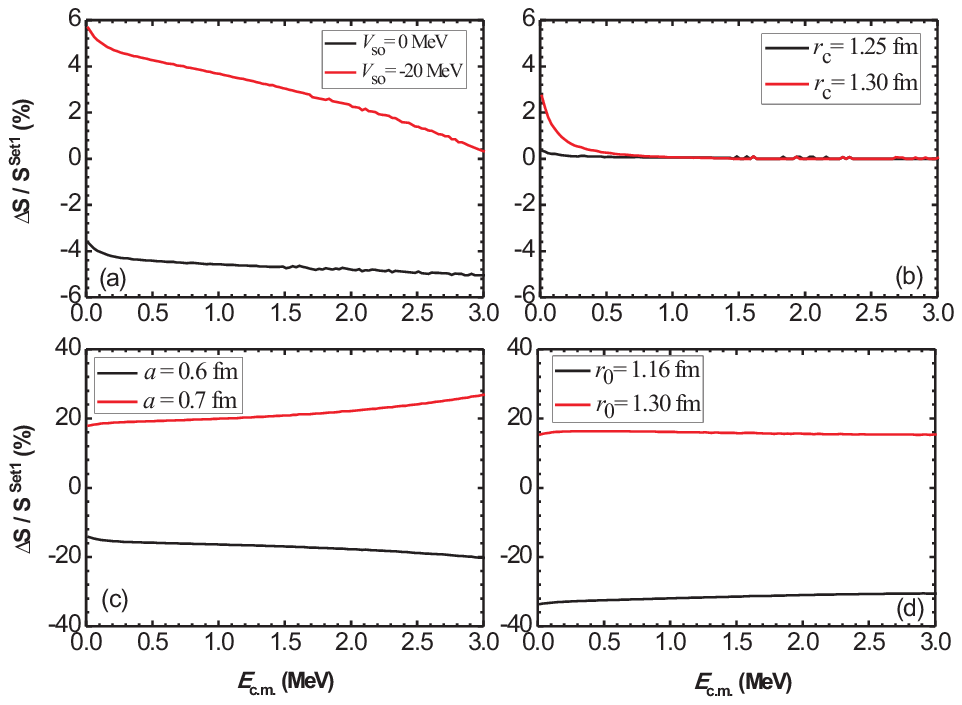}
\vspace{-2mm}
\caption{\label{fig5} (Color online) Dependence of $S_\mathrm{dc}$ factors on parameters (a) spin-orbit potential $V_{so}$,
(b) Coulomb radius parameter $r_c$, (c) diffuseness $a$, and (d) optical-model (real) potential radius parameter $r_0$. Here, radius
$r$ is defined as $r$=$R$/$A^{\frac{1}{3}}$, with $r_0$=$r_{so}$, $a_0$=$a_{so}$ in all calculations.}
\end{center}
\end{figure*}

In addition, the parameter dependence on $S_\mathrm{dc}(E)$ has been studied and the results are shown in Fig.~\ref{fig5}.
It shows that $S_\mathrm{dc}$ factor is insensitive to the parameters $V_{so}$ and $R_c$ (or $r_c$), but rather sensitive to the
parameters $R_0$ (or $r_0$) and $a$. The choice of parameter ranges is based on the literature values~\cite{bib:per63,bib:var91,bib:ili97}.
The error of the present DC rate is estimated simply by adding in quadrature the uncertainties originating from the potential parameters
discussed above; it is about $\sim$40\% in the energy range of 0--3 MeV. The DC rate as a function of temperature is
calculated by numerical integration of our calculated $S$ factors using an EXP2RATE code~\cite{bib:exp2}.


\end{document}